%% file: paper.tex
\documentclass[10pt,conference]{IEEEtran}
\IEEEoverridecommandlockouts
\input{macros}

\usepackage{booktabs}
\usepackage{caption}
\usepackage{subcaption}
\usepackage{cite}
\usepackage{amsmath,amssymb,amsfonts}
\usepackage{algorithmic}
\usepackage{graphicx}
\usepackage{textcomp}
\usepackage{xcolor}
\def\BibTeX{{\rm B\kern-.05em{\sc i\kern-.025em b}\kern-.08em
    T\kern-.1667em\lower.7ex\hbox{E}\kern-.125emX}}
\begin{document}

\title{CoCoFuzzing: Testing Neural \underline{Co}de Models with \underline{Co}verage-Guided \underline{Fuzzing}
}
\author{Anonymous}
\author{
    \IEEEauthorblockN{Moshi Wei\IEEEauthorrefmark{1}, Yuchao Huang\IEEEauthorrefmark{2}, Jinqiu Yang\IEEEauthorrefmark{3}, Junjie Wang\IEEEauthorrefmark{2},
    Song Wang\IEEEauthorrefmark{1}}
    \IEEEauthorblockA{\IEEEauthorrefmark{1}York University, Canada
    \\\{moshiwei, wangsong\}@yorku.ca}
    \IEEEauthorblockA{\IEEEauthorrefmark{3}Concordia University, Canada
    \\\{jinqiuy\}@encs.concordia.ca}
     \IEEEauthorblockA{\IEEEauthorrefmark{2}Institute of Software, Chinese Academy of Sciences, China	
    \\\{hycsoge,junjie\}@iscas.ac.cn}
}

\maketitle

\begin{abstract}
\input{sec/abstract}
\end{abstract}


\input{sec/introduction}

\input{sec/motivation}
\input{sec/approach}
\input{sec/setup}

\input{sec/result}
\input{sec/discussion}
\input{sec/related}
\input{sec/conclusion}

\bibliographystyle{IEEEtran}
\bibliography{sample-base}

\end{document}

%% file: macros.tex
\usepackage{soul}
\usepackage{hyperref}
\usepackage{multirow}
\usepackage{listings}
\usepackage{fancybox}
\usepackage{enumitem}
\usepackage{graphicx}
\usepackage{color}
\usepackage{xcolor}
\usepackage{array}
\usepackage{amsmath}
\usepackage{xspace}
\usepackage{url}
\usepackage{tikz}
\usepackage{caption}
\usepackage{pgfplots}
\pgfplotsset{compat=1.16}
\usepackage{pgf-pie}
\usetikzlibrary{pgfplots.statistics,calc}
\usepackage{algorithm}
\usepackage{algorithmic}

\usepackage{tcolorbox}
\makeatletter
\newcommand{\mybox}[1]{%
	\setbox0=\hbox{#1}%
	\setlength{\@tempdima}{\dimexpr\wd0+13pt}%
	\begin{tcolorbox}[boxrule=0.5pt, colback=white, arc=4pt,
		left=6pt,right=6pt,top=6pt,bottom=6pt,boxsep=0pt]
		#1
	\end{tcolorbox}
}

\definecolor{songcolor}{RGB}{191,191,191}
\newcommand{\todoc}[2]{{\textcolor{#1}{\textbf{#2}}}}

\newcommand{\todoorange}[1]{\todoc{orange}{\textbf{[[#1]]}}}

\newcommand{\song}[1]{\todoorange{Song: #1}}

\newcommand{\tool}{\texttt{CoCoFuzzing}}
\newcommand{\neuralCodeSum}{{NeuralCodeSum}}
\newcommand{\neuralcodesum}{{NeuralCodeSum}}
\newcommand{\codeseq}{{CODE2SEQ}}
\newcommand{\codevec}{{CODE2VEC}}


%% file: sec/abstract.tex
{Deep learning-based {code processing} models have shown good performance for tasks such as predicting method names, summarizing programs, and comment generation.} 
However, despite the tremendous progress, deep learning models are often prone to adversarial attacks, which can significantly threaten the robustness and generalizability of these models by leading them to misclassification with unexpected inputs. To address the above issue, many deep learning testing approaches have been proposed, however, these approaches mainly focus on testing deep learning applications in the domains of image, audio, and text analysis, etc., which cannot {be directly applied} to {neural models for code} due to the unique properties of programs.

In this paper, we propose a coverage-based fuzzing framework, {\tool}, for testing deep learning-based code {processing} models. 
In particular, we first propose ten mutation operators to automatically generate valid and semantically preserving source code examples as tests; 
then we propose a neuron coverage-based approach to guide the generation of tests. 
We investigate the performance of {\tool} on three state-of-the-art neural code models, i.e., {\neuralCodeSum}, {\codeseq}, and {\codevec}. 
Our experiment results demonstrate that {\tool} can generate valid and semantically preserving source code examples for testing the robustness and generalizability of these models and improve the neuron coverage.  
Moreover, these tests can be used to improve the performance of the target neural code models through adversarial retraining.


%% file: sec/introduction.tex
\section{Introduction}
\label{sec:intro}

Deep learning (DL) has recently been successfully applied to accelerate many tasks in automated source code processing 
such as prediction of variable names~\cite{alon2018code2seq,alon2019code2vec}, code summarization~\cite{ahmad2020transformer,hu2018deep,liang2018automatic,allamanis2016convolutional}, and comment generation~\cite{hu2020deep,chen2019neural}. 
Most of these deep learning-based code models (i.e., neural code models for short) have reported good performance.  


However, deep learning models are widely known to suffer from adversarial attacks~\cite{ma2018deepgauge,pei2017deepxplore}, i.e., a subtly-modified input can lead neural networks to misclassifications and result in severe erroneous behavior of DL 
models. 
More thorough testing of neural networks can improve their reliability and robustness. 
Many deep learning testing approaches have been proposed~\cite{xie2019deephunter,tian2018deeptest,zhang2019cagfuzz,alzantot2018generating}, however, most of {these} 
approaches mainly focus on generating tests for deep learning models in the domains of image, audio, and text analysis by using mutations such as image adjustment, image scaling, image rotation, and noise addition, etc., which cannot be directly applied to {neural code models}. 
In addition, different from adversarial example generation for image, audio, and natural languages, the structured nature of programming languages brings new challenges, i.e., program must strictly follow the rigid lexical, grammatical and syntactical constraints, and tests generated for a program are expected to fully preserve the original program semantics. 

Recently, some adversarial example generation approaches for source code were proposed, e.g., \cite{zhang2020generating} performed variable name replacements to perturb the programs and \cite{yefet2019adversarial} proposed to use both variables renaming and dead code (i.e., unused variable declaration) insertion to generate semantically equivalent adversarial examples. {However, the types of perturbations introduced by the two transformations are limited, as we have witnessed many other types of noise/perturbation in real-world software programs that sharing the same semantics~\cite{allamanis2019adverse}.  Hence only using the two operators may miss numerous corner cases for the purpose of thoroughly testing neural code models.}


In this paper, we propose a coverage-based fuzzing framework, {\tool}, to test neural code models.  
Specifically, we first propose and implement ten mutation operators, which represent various real-world semantic preserving transformations of programs, to automatically generate valid and semantically preserving source code examples as tests.  
Then we utilize a neuron coverage based
guidance mechanism for systemically exploring
different types of program transformations and {guiding} the generation of tests. 
We investigate the performance of {\tool} on three state-of-the-art typical neural code models, i.e., {\neuralCodeSum}~\cite{ahmad2020transformer} (leverages a self-attention-based neural network to generate summarization of programs), {\codeseq}~\cite{alon2018code2seq} (builds a AST-based RNN neural network to predict method names), and {\codevec}~\cite{alon2019code2vec} (uses a path-based attention model for learning code embeddings to represent a method).  

Our experiment results demonstrate that {\tool} can generate valid and semantically preserving tests for examining the robustness and generalizability of neural code models. Specifically, the newly-generated tests by {\tool} can reduce the performance of {\neuralCodeSum}, {\codeseq}, and {\codevec} by 84.81\%, 22.06\%, and 27.58\% respectively. 
Moreover, we find that these new tests by {\tool} can also be used to improve the performance of the target neural code models through adversarial retraining. 
Specifically, performance can be improved 35.15\% on {\neuralCodeSum}, 8.83\% on {\codeseq}, and 34.14\% on {\codevec} by retraining the models with synthetic data generated by {\tool}. 

This paper makes the following contributions:
\begin{itemize}
\item We propose a coverage-based fuzzing framework {\tool} for testing the robustness of neural code models. 
To the best of our knowledge, {\tool} is the first fuzzing framework for testing neural code models. We have released the implementation of our tool to facilitate the replication of this  study\footnote{https://doi.org/10.5281/zenodo.4000441}.

\item We implement ten mutation operators, which represent various real-world semantic preserving transformations of programs, to automatically generate tests. 

\item We utilize and experiment a neuron coverage-based guidance mechanism for systemically exploring the large search space constituted of different types of program transformations and measure the adequacy of the fuzzing. 

\item We evaluate {\tool} on three state-of-the-art neural code process models, i.e., {\neuralCodeSum}, {\codeseq}, and {\codevec}. The experiment results show the effectiveness of our tool. 

\end{itemize}	 
The rest of this paper is organized as follows. 
Section~\ref{sec:bg} presents
the background of neural code models.   
Section~\ref{sec:approach} describes the
methodology of our proposed {\tool}. 
Section~\ref{sec:experiment} shows the setup of our experiments. 
Section~\ref{sec:result} presents the result of our study.  Section~\ref{sec:discussion} discusses the threats to the validity of this work. Section~\ref{sec:related} presents related studies. Section~\ref{sec:conclusion} concludes this paper.


%% file: sec/motivation.tex
\section{Background}
\label{sec:bg}

\subsection{Neural Code Models}
\label{sec:bg1}
The growing availability of open-source repositories creates new opportunities for using deep learning to accelerate code processing tasks 
such as prediction of variable names~\cite{alon2018code2seq,alon2019code2vec,alon2018general}, code summarization~\cite{ahmad2020transformer,hu2018deep,liang2018automatic,allamanis2016convolutional}, and comment generation~\cite{hu2020deep,chen2019neural}.
The deep neural networks used in neural code processing models mainly can be categorized into two types: (1) Recurrent Neural Network (RNN) and (2) Attention Neural Network. 
We provide a brief description of each architecture below.

\noindent \textbf{RNN architecture.}
A recurrent neural network is a neural network that consists of a hidden state $h$ and an optional output $y$ which operates on a variable-length sequence $x=(x_{1}, ... , x_{T})$. And at each time step $t$, the hidden state $h_{t}$ of the RNN is updated
by $h_{t}=f(h_{t-1}, x_{t})$. $f$ is a non-linear activation function. Standard RNNs are not capable of learning ``long-term dependencies'', i.e., they may not propagate information that appeared earlier in the input sequence later because of the vanishing and exploding gradient problems. 
Long short-term memory LSTM~\cite{hochreiter1997long} has been proposed to address the above issue. It introduces additional internal states, called memory cells, that do not suffer from the vanishing gradients and it controls what information will be propagated. 

\noindent \textbf{Attention network architecture.}
Attention neural models have been recently widely used in the field of natural language process and have achieved very promising results~\cite{yin2016abcnn}. 
A neural attention mechanism equips a neural network with the ability to focus on a subset of its inputs when processing a large amount of information. Based on the network characteristics, attention neural networks have three main variants, i.e., global and local attention~\cite{luong2015effective}, hard and soft attention~\cite{shen2018reinforced}, and self-attention~\cite{zhang2019self}. 
Recently, Google proposed the Transformer model~\cite{vaswani2017attention} for natural language process tasks, which is the first transduction model relying entirely on self-attention to compute representations of its input and output. 

\subsection{Fuzz Testing}
\label{sec:bg2}
Software fuzzing in short generates mutants by modifying valid seed inputs to test a program~\cite{zalewski2014american}. The mutants are failed tests if they cause abnormal behaviors (e.g., crash the system under test) otherwise are passed tests.  
Coverage-guided fuzzing has been proposed and widely used to find many serious bugs in real software~\cite{aizatsky2016announcing}, in which a fuzzing process maintains an input data corpus for the program under consideration. 
Changes are made to those inputs according to some mutation procedure, and mutated inputs are kept in the corpus when they exercise new ``coverage''. 
Fuzzing testing techniques of traditional software leverage code coverage metrics that track which
lines of code have been executed and which branches have been taken~\cite{serebryanylibfuzzer,zalewski2014american}. 
However, these traditional fuzzing frameworks could not be directly applied to deep neural network-based
software due to the fundamental difference in the programming paradigm and the development process~\cite{xie2019deephunter}, i.e., 
a neural network run on different inputs will often execute the same lines of code and take the same branches, yet can produce significantly different behavior due to the difference in input values. 

To test deep neural networks, recently many coverage-guided fuzz testing frameworks have been proposed~\cite{xie2019deephunter,odena2019tensorfuzz,tian2018deeptest}, which applies the neuron coverage metrics to guide the fuzzing test to deep learning applications in the domain of image processing and they perform image transformations such as blurring and shearing to generate fuzzing inputs from seed images.  

%% file: sec/approach.tex
\section{The Approach of \tool}
\label{sec:approach}


\input{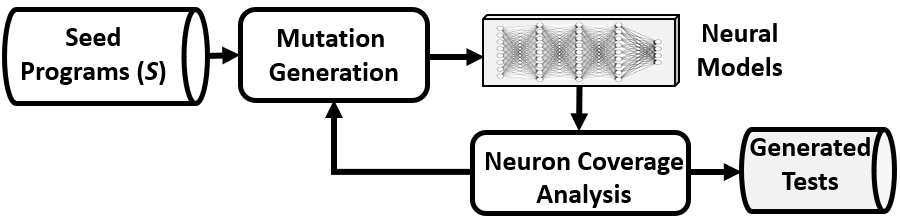}
In this section, we describe the approach of {\tool}.
Figure~\ref{fig:overview} shows the overview and Algorithm~\ref{alg:algorithm0} describes the main algorithm in detail. 
{\tool} takes a set of initial seed programs and a neural code model as the input, and produces new test sets iteratively through mutation generation (Section~\ref{sec:step1}) and neuron coverage analysis (Section~\ref{sec:step2}).

\subsection{Overview of {\tool}}
\label{sec:overall}
\input{sec/algorithm2}
Algorithm~\ref{alg:algorithm0} shows how {\tool} works step by step.
{\tool} begins by selecting one seed program from
the seed queue (line 1--2), then performs source code transformations to the input based on a list of pre-defined mutation operators (line 11). Given the non-trivial search space constructed by combinations of mutation operators, {\tool} employs a neuron coverage-guided (short for NC-guided) approach (line 5--25), to search for certain types of mutated programs. At a high level, {\tool} searches for program mutations that activate new neurons (line 13--17) while controlling the maximum mutations on one seed program (line 5 and a threshold \textit{MAX}), for both naturalness and the feasibility of utilizing metamorphic testing (i.e., mutated programs can use the same test oracle as the seed program). 

Note that in traditional coverage-guided fuzzers for computer programs~\cite{serebryanylibfuzzer,zalewski2014american}, a seed can be reused and mutated for multiple times, e.g., these tools often use either time constraints or the number of generated mutants as the termination condition and a single seed can generate thousands of mutants~\cite{rebert2014optimizing}. 
However, a similar process cannot be directly applied for testing neural code models as there are no explicit oracles (i.e., program crashes) available to assess the result of a seed program mutated multiple times~\cite{xie2019deephunter}. To bypass the oracle challenge in testing neural code models, we leverage metamorphic testing, similar to \cite{xie2019deephunter,tian2018deeptest}. A mutated program shares the same oracle as the seed program if the transformation is semantic-preserving. 
Note that mutating a program could hurt its naturalness as the inserted/mutated code could be noise to the original program.  
Hence, we use a threshold $MAX$ to limit the maximum number of mutations applied on a seed program, which means a seed program at most can be mutated for $MAX$ times accumulatively. 
{Note that the design of $MAX$ controller does not guarantee the naturalness of the inserted mutation code. A robust code model should have the ability of ignoring the inserted noise code despite its naturalness.} 

To find an appropriate value for $MAX$, we randomly selected 1,000 sample programs from the test datasets of the three studied neural code models (details are in Section~\ref{sec:sub}). For each sample, we randomly select $MAX$ mutation operators and apply them to the given sample to generate mutants.  
We experiment $MAX$ with values from 1 to 10. 
{In the most conservtive senario, we assume the generated code are entirely un-natural}.Hence, We measure the naturalness of the mutated program by using the percentage of the generated code.

\input{figure/dis_max}
Figure~\ref{fig:dis_max} shows the average percentage of generated code against the seed program under different $MAX$. 
With the increass of $MAX$, the naturalness of code decreases dramatically. For example, when $MAX$ is equal to 1, on average only 9.25\% code in a mutated program is noise (i.e., generated code), while $MAX$ increases to 10, the percentage of noise code is 59.60\% on average. 

Prior work showed that overall there is about 28\% noise code in real-world software projects~\cite{eder2012much}. 
Thus, for all the subsequent experiments in this paper, 
we set $MAX$ to three, which yields less 30\% noise code on average. 
\subsection{Neuron Coverage Analysis}
\label{sec:step2}
Algorithm~\ref{alg:algorithm0} details how {\tool} uses neuron coverage in searching for new test data (line 5--25). In each iteration (line 10--18), {\tool} tries all the mutation operators and identifies one mutated program that activates the most number of new neurons. This mutated program is produced as one test data (line 24) and also kept for the next mutation iteration. Controlled by the threshold \textit{MAX} (line 5), this mutant might be continuously mutated in the next iteration, until the number of mutations applied per original seed program reaches the threshold, which is empirically set to three in our experiments.

Two methods \textit{neuronActivation} and \textit{newNeurons} in Algorithm~\ref{alg:algorithm0} are for neuron coverage analysis. 
The \textit{neuronActivation} method takes a trained neural model and one input and produces a set of activated neurons as output. {For example, given a trained model in {PyTorch}}, we inject listeners to monitor the neurons. 
Then, we run the inference or prediction of the trained model with the input and collected the output of each neurons using the injected listeners. 
Similar to prior work~\cite{tian2018deeptest,xie2019deephunter}, for each neuron, we scale the value of the neuron to the range of 0 to 1 and compare the scaled value to a threshold. 
If the scaled output is greater than the threshold (i.e., we use 0.4, the same value with \cite{tian2018deeptest,xie2019deephunter}), we consider the neuron activated, otherwise inactivated. 

The \textit{newNeurons} method is very similar to \textit{neuronActivation}, which takes a model, one input data, and a set of activated neurons (i.e., $NC_p$ in line 12) as the input. The method \textit{newNeurons} calculates the activated neurons by the input, and outputs a set of newly activated neurons by the input but not $NC_p$. 
The implementation of neuron coverage analysis in {\tool} is model-independent and can be widely applied to various deep learning models.

\subsection{Mutation Generation}
\label{sec:step1}
In \tool, we adopt and implement specialized mutation operators for mutating programs. To bypass the oracle problem, our proposed mutation operators preserve program semantics. Hence, the mutated programs should have the same test oracles with the original programs, i.e., metamorphic testing.
Prior studies~\cite{tian2018deeptest, xie2019deephunter} in testing neural models propose mutation operators for inputs such as image and audio, e.g., image scaling, image rotation, and noise addition, etc. These operators cannot be directly applied to source code programs as programs must strictly follow the rigid lexical, grammatical, and syntactical constraints. 

In this work, we propose to use a set of ten mutation operators that can be applied to method level to generate
semantically-equivalent methods. The mutation operations include transformations
ranging from common refactoring like variable renaming to more intrusive ones like adding unreachable branches. Table~\ref{tab:opts} summarizes the ten mutation operators and how each mutation operator can be applied to transform a program. 
Note that, among the ten operators, \textit{Op1} and \textit{Op10} are proposed and experimented  in prior  studies~\cite{yefet2019adversarial,zhang2020generating} to generate adversarial examples for neural code models. The other eight mutation operators are first used in this work. These mutation operators target at a diverse types of program transformations.

We use the open-source java parser package javalang\footnote{\url{https://pypi.org/project/javalang/}} 
for code parsing and tokenization. 
Given a program $T$ and a mutation operator $O$, 
we first use the AST parser from javalang to convert $T$ to a list of sub-AST where each element is the AST representation of a statement. 
We then iterate the sub-AST list and mark the possible positions for $O$ regarding its specific transformation requirements. The detailed process of mutation generation for each operator is as follows.



\begin{itemize}
\item \textbf{(Op1) Dead store~\cite{yefet2019adversarial}:} inserts an unused variable declaration with one primitive type (e.g., string, int, double, and long, etc.) to a randomly selected basic block in the program.  
The name of the variable is a string of eight characters randomly generated in the form of [a-z]. 
Only one dead store is added in each transformation by this operator.

\item \textbf{(Op2 and Op3) Obfuscating:} rewrites a numerical value or variable and its usages in a statement by adding and deleting the same random numerical value of the same type. For example, $x = 1.0;$ can be mutated to ${x = 1.0 + 0.1 -0.1;}$ or $x = 1.0+0-0;$. If one program contains more than one numerical variables, we randomly pick one to perform the transformation. This operator only works on assignment, declaration, or return statements.

\item \textbf{(Op4) Duplication:} duplicates a randomly selected assignment statement and insert immediately after its current location. To avoid side-effect, the applicable assignment statement is limited to the ones without using method invocation.

\item \textbf{(Op5 to Op9) Unreachable loops/branches:} inserts an unreachable loop or branch (including \textit{if} statement, \textit{for} statement, \textit{while} statement, and \textit{switch} statement) into a randomly selected basic block in the program. 
The condition of the inserted loop or branch is always false to make it unreachable. 

\item \textbf{(Op10) Renaming~\cite{yefet2019adversarial,zhang2020generating}:} renames a local variable declared in a program. If there exist multiple variables, we randomly select one for the mutation. The new name of the variable will be in the form of [a-z]. 
\end{itemize}

\input{table/mutationOpts}


Note that some of the proposed operators, i.e., \textit{Op1} and \textit{Op5}--\textit{Op9}, can be inserted into any locations in a given program. In this work we randomly pick a location to apply these mutation operators because through experiments we find that the chosen locations are not correlated with the effectiveness of these mutation operators (details are in Section~\ref{sec:dis1}).

%% file: figure/overall.tex
\begin{figure}[t!]
\centering
\includegraphics[width=1\columnwidth]{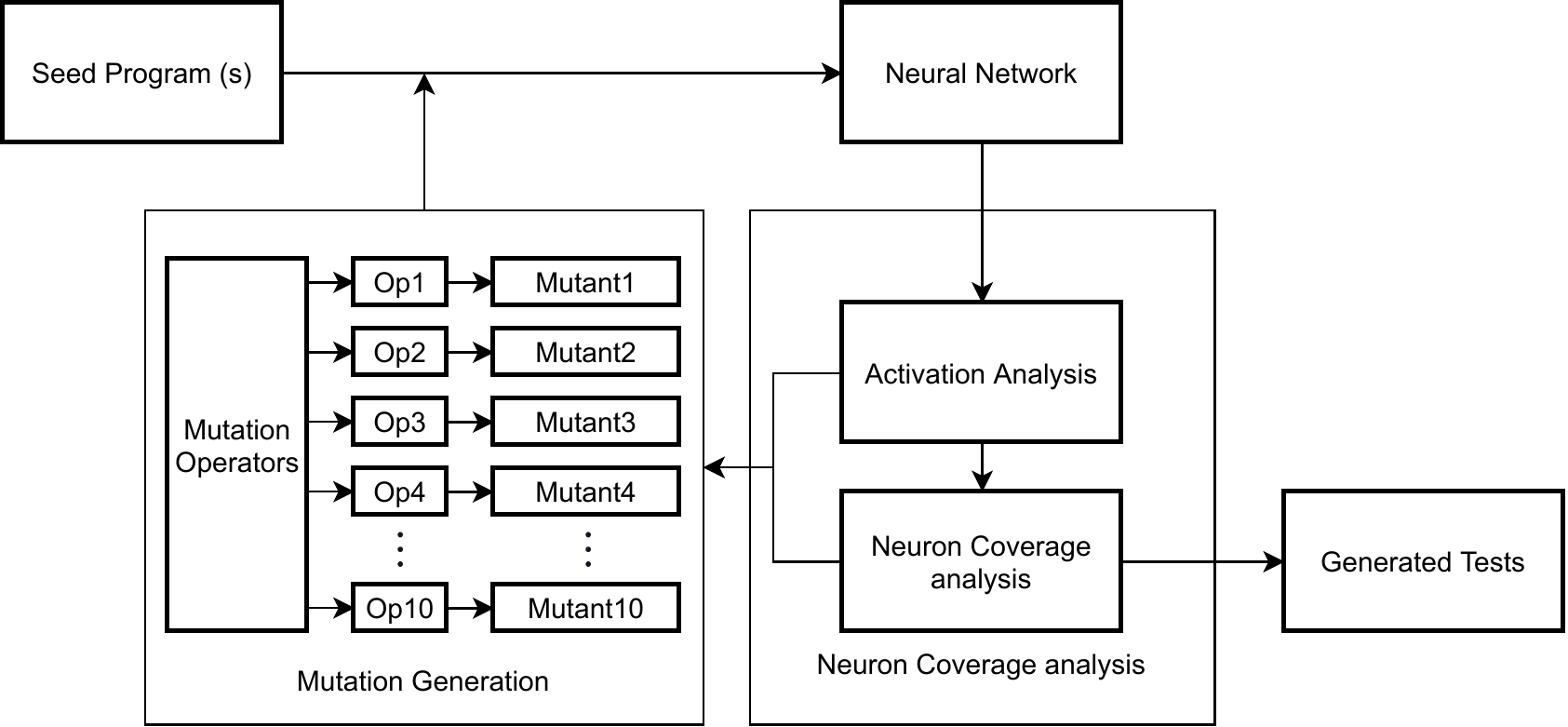}
\caption{The overview of our proposed {\tool}.}
\label{fig:overview}
\end{figure}

%% file: sec/algorithm2.tex
\renewcommand{\algorithmicrequire}{}
\renewcommand{\algorithmicensure}{}
\begin{algorithm}[t]
\caption{\small Coverage-Guided Test Generation in {\tool} \label{alg:algorithm0}}
\hrule
	\begin{algorithmic}[1]
		\small
		\vspace{.2cm}
		\REQUIRE 
		\textbf{Input:}
		seed programs \textit{S}; 
		target neural code model \textit{NM};\\
		mutation operators \textit{Ops};
		Maximum number of mutations \textit{MAX}; \\
        \textbf{Variables:} sets of activated neurons $NC_p$, $NC_{curr}$, \textit{bestActivationSet}; \\
		\ENSURE\textbf{Output:}
		 generated tests \textit{T};
		\WHILE {$S$ is not empty}
		\STATE p $\leftarrow$ S.pop()
		\STATE $NC_p$ $\leftarrow$ neuronActivation(p, NM) 
        \STATE numTries = 1
		\WHILE {numTries $<=$ MAX}
		\STATE numTries++
		\STATE bestMutant $\leftarrow$ null 
		\STATE bestActivationSet $\leftarrow$ $\emptyset$
		\STATE bestActivationCount $\leftarrow$ 0
		\FOR   {op in Ops } 
		\STATE currMutant = apply(p, op)
\STATE \small{$NC_{curr}$ $\leftarrow$ newNeurons(currMutant, $NC_p$, NM)}
		\IF  { size($NC_{curr}$) $>$ bestActivationCount }
		\STATE bestActivationCount $\leftarrow$ size($NC_{curr}$)
		\STATE bestActivationSet $\leftarrow$ $NC_{curr}$
		\STATE bestMutant $\leftarrow$ currMutant
		\ENDIF 
		\ENDFOR
		\IF  { bestMutant == null }
		\STATE break
		\ENDIF
		\STATE p $\leftarrow$ bestMutant
		\STATE $NC_p$ $\leftarrow$ bestActivationSet $\cup NC_p$
		\STATE T.push(p) 
		\ENDWHILE
		\ENDWHILE
	\end{algorithmic}
	\hrule
\end{algorithm}


%% file: figure/dis_max.tex
\begin{figure}[t!]
\centering
\footnotesize
	\begin{tikzpicture}[thick, scale=0.8]
	\begin{axis}[
	ybar,
	height=3.50cm,
	width= 10cm,
	enlargelimits=0.1,
	ymin=0,
	ylabel={Percentage},
	symbolic x coords={1, 2, 3, 4, 5, 6, 7, 8, 9, 10},
	xtick=data,
	x tick label style={font=\footnotesize},
	]
\addplot+[ybar,gray,  fill=gray]coordinates{
(1,  0.0925) 
(2,  0.2103) 
(3,  0.3000) 
(4,  0.3681) 
(5,  0.4250) 
(6,  0.4735) 
(7,  0.5128) 
(8,  0.5447) 
(9,  0.5731) 
(10, 0.5965)};
\end{axis}
\end{tikzpicture}
\caption{\small Average percentages of noise code in the mutants generated with different $MAX$ (i.e., X axis).}
\label{fig:dis_max}
\end{figure}
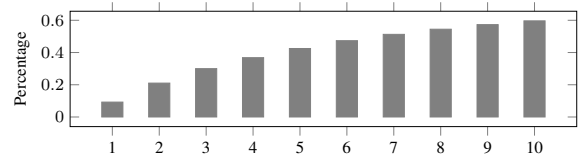

%% file: table/mutationOpts.tex
\begin{table}[t!]
\centering
\caption{Ten Semantic-Preserving Mutation Operators Applied in {\tool}.}
\label{tab:opts}
\setlength{\tabcolsep}{3pt}
\scalebox{0.8}{
\begin{tabular}{lll}
\toprule \hline
\textbf{NO.} & \textbf{Operator name}  & \textbf{Description}  \\ \hline
Op1~\cite{yefet2019adversarial} & dead store            & Inserting unused variable declarations\\ \hline

Op2 & numerical obfuscating & \begin{tabular}[c]{@{}l@{}}Obfuscating the numerical variables via\\ adding/deleting a same numerical value\end{tabular} \\ \hline
Op3 &adding zero & \begin{tabular}[c]{@{}l@{}}Obfuscating the numerical values via\\ adding zero \end{tabular} \\ \hline

Op4 & duplication          & Duplicating assignment statements\\ \hline
Op5 & unreachable \textit{if}   & Inserting unreachable \textit{if} statements \\ 
Op6 & unreachable \textit{if-else}  & Inserting unreachable \textit{if-else} statements \\ 
Op7 & unreachable \textit{switch}  & Inserting unreachable  \textit{switch} statements \\ 
Op8 & unreachable \textit{for}      & Inserting unreachable  \textit{for} statements \\ 
Op9 & unreachable \textit{while}   & Inserting unreachable  \textit{while} statements \\ \hline
Op10~\cite{yefet2019adversarial,zhang2020generating} & renaming              & Renaming user-defined variables \\ \bottomrule \hline
\end{tabular}
}
\end{table}

%% file: sec/setup.tex
\section{Experiment Setup}
\label{sec:experiment}
We run all the experiments on the Google Cloud Computing Platform. Specifically, we use two n1-highmem-2 virtual central processing unit (vCPU) with 13 gigabyte memory in total and one NVIDIA Tesla T4 GPU as hardware of our experiment machine. 
We use PyTorch 1.4 for running {\neuralCodeSum}, TensorFlow 1.15 for {\codeseq}, and TensorFlow 2.1 for {\codevec}. 

\subsection{Subject Models and Datasets} 
\label{sec:sub}

\subsubsection{Studied Models}
\label{sec:smodels}

In this work, to evaluate {\tool}, we use three state-of-the-art neural code models that adopt different neural network characteristics, i.e., {\neuralCodeSum}~\cite{ahmad2020transformer},  {\codeseq}~\cite{alon2018code2seq}, and {\codevec}~\cite{alon2019code2vec}. 
For our experiments, we reuse the pre-trained models of the three studied neural code models, which are released in their papers.
{All the three models are common in learning code embedding from large-scale data and generating a high-level summary given a method. However, the three models are different in many aspects. First, they utilize different model architectures. Second, they have different representations of code. Third, the level of details in the generated summaries from the three models is different, i.e., {\neuralCodeSum} generates full English sentence as output while {\codeseq} and {\codevec} produce  the name of the method.} 
We briefly describe the details of the three models below.


\noindent \textbf{\neuralCodeSum}~\cite{ahmad2020transformer} uses  Transformer~\cite{ahmad2020transformer} (consists of stacked multi-head attention and parameterized linear transformation layers for both the encoder and decoder) to generate a natural language summary
given a piece of source code. Both the code and
the summary is a sequence of tokens that are represented by a sequence of vectors. To allow the Transformer to utilize the order information of source code tokens, {\neuralCodeSum} encodes both the absolute position and pairwise relationship of source code tokens.

\noindent \textbf{\codeseq}~\cite{alon2018code2seq} uses an encoder-decoder architecture to encode paths node-by-node and generate label as sequences. In {\codeseq}, the encoder represents a method’s body as a set of AST paths where each path is compressed to a fixed-length vector using a bi-directional LSTM which encodes paths node-by-node. 
The decoder uses attention to select relevant paths while decoding and predicts sub-tokens of target sequence at each step when generating the method’s name. 

\noindent \textbf{\codevec}~\cite{alon2019code2vec} 
proposes a path-based attention model for learning vectors for arbitrary-sized snippet of code. The model allows to embed a program into a continuous space.
Specifically it first extracts syntactic paths from within a code snippet, i.e., ASTs, and then it represents them as a bag of distributed vector representations.
An attention network will be used to compute a learned weighted average of the path vectors in order to produce a single code vector. 

\subsubsection{Dataset}
\label{sec:sdata}
\input{table/dataset}

We perform our experiments using the original datasets associated with each of the three studied neural code models. 
Table~\ref{tab:data} lists the basic statistics of the three datasets. Specifically, the dataset of {\neuralCodeSum} contains 9.7K open source Java projects hosted in GitHub and each of the projects has at least 20 stars. 87.1k methods that have JavaDoc comments were collected. 
{\codeseq} created three Java datasets, which are different in size, the popularity of the open-source software where the data is from, and the distribution of training, validation and testing. Among the three Java datasets by {\codeseq}, we decided to use one, i.e., \textit{Java-small}. Despite its relatively small size among the three, it is commonly used by prior studies~\cite{allamanis2016convolutional, alon2019code2vec}, including {\codevec}. Also, all the projects in \textit{Java-small} are large-scale and mature open-source software, in comparison with the other two datasets (i.e., many are from smaller and less popular projects). 
\textit{Java-small} contains 11 relatively large Java projects 9 for training, 1 for validation, and 1 for testing. Overall, it contains about 772.9K methods. 

\subsection{Evaluation Metrics}
\label{sec:metrics}
We use the same evaluation metrics with the original papers of {\neuralCodeSum}, {\codeseq}, and {\codevec} for comparison purpose.
{\neuralCodeSum} 
adopts BLEU~\cite{papineni2002bleu} to evaluate its performance on predicting the summarization of programs. BLEU is widely used to assess the quality of machine translation systems~\cite{cho2014learning}.  
BLEU's output is always a number between 0 and 1. This value indicates how similar the predicting summarization is to the ground-truth, with values closer to 1 representing more similar. 

Differently, {\codevec} and {\codeseq} used precision, recall, and F1 for measuring performance.
Precision and Recall are computed in a per-subtoken basis. F1 is the weighted average of Precision and Recall. 


\subsection{Baseline}
\label{sec:baselines}
\subsubsection{Baseline for Mutation Operators} 

As two of the ten mutation operators are used in prior studies~\cite{yefet2019adversarial,zhang2020generating}, namely \textit{Op1} (i.e., dead code inserting) and \textit{Op10} (i.e.,renaming). 
We treat them as baselines to evaluate the eight new mutation operators, i.e., \textit{Op2}-- \textit{Op9}.  

\subsubsection{Baseline for NC-Guided Test Generation} 
{\tool} uses neuron coverage to guide the test generation via the combination of mutation operators. 
To evaluate the effectiveness of {\tool} we have also designed a baseline approach that generates tests without neuron coverage guidance, i.e., \textit{Random@K}, which randomly picks $K$ mutation operators to generate mutants as tests.  
$K$ indicates the maximum number of mutations tries on a seed program. As we described in Section~\ref{sec:approach}, {\tool} limits the maximum number of mutation tries to three, thus in our experiment we also set $K$ to three. 
Given a seed program, \textit{Random@3} 
first randomly selects a mutation operator $op_{i}$ and mutate the sample to get the mutant $M1$. 
Then, it mutates $M1$ with another randomly selected mutation operator $op_{j}$ to get a new mutant $M2$, after that it further mutates $M2$ with a randomly selected mutation operator $op_{h}$ to generate a new mutant M3. Through the process, $op_{i}$, $op_{j}$, and $op_{h}$ can be the same operator.
\textit{Random@3} produces a total of 3 mutants for a given seed program.  

\subsection{Research Questions}
\label{sec:rqs} 
We have implemented {\tool} as a self-contained fuzz testing framework in Python based on deep learning framework Keras, TensorFlow, and Pytorch. With {\tool}, we perform a large-scale comparative study to answer the following four research questions.

\vspace{4pt}
\noindent  \textbf{RQ1:} \textit{Are the neural code models robust against simple perturbations?}

Robustness has been extensively studied in classic deep learning application domains, e.g., image processing, speech recognition, and NLP. Many of these deep learning models have been shown that they could be easily attacked by simple perturbations. As the first study on exploring the robustness of neural code models, in this RQ, we explore whether they also suffer similar issues.
 
\vspace{4pt}
\noindent  \textbf{RQ2:} \textit{What is the effectiveness of each mutation operator?}

This RQ illustrates the effectiveness of each mutation operator (a total of ten in Table~\ref{tab:opts}) regarding its capability of introducing perturbations that can affect the performance of neural code models and activating different neurons. 
Furthermore, this RQ also examines the effectiveness of the two baseline operators (i.e., \textit{Op1} and \textit{Op10}) in comparison to other eight new mutation operators. 

\vspace{4pt}
\noindent  \textbf{RQ3:} \textit{What is the effectiveness of the NC-guided test generation in {\tool}?}

 {\tool} uses neuron coverage to guide the search for new tests that are continuously transformed through multiple mutation operators. This RQ explores the performance of {\tool} and compares it with our constructed baseline, i.e., \textit{Random@3}.


\vspace{4pt}
\noindent  \textbf{RQ4:} \textit{Is {\tool} useful for improving neural code models?}

This RQ explores whether the synthetic programs can improve the neural code models, i.e., whether retraining with  with \tool's synthetic programs can make these models more robust. 

%% file: table/dataset.tex
\begin{table}[t!]
\centering
\caption{ Experimental datasets. `Pro' is the number of projects.}
\label{tab:data}
\scalebox{0.8}{
\begin{tabular}{lcccccc}
\toprule \hline
\multicolumn{1}{c}{Model}       & Language   & \multicolumn{1}{l}{\#Pro} &  \multicolumn{1}{l}{\#Training}& \multicolumn{1}{l}{\#Validation}& \multicolumn{1}{l}{\#Test}& \multicolumn{1}{l}{\#Method} \\ \hline
{\neuralCodeSum} & Java            & 9.7k     &  69.7k      &    8.7k    &         8.7k           & 87.1K                         \\ \hline
{\codeseq}       & \multirow{2}{*}{Java}  & \multirow{2}{*}{11}     & \multirow{2}{*}{692.0k} &\multirow{2}{*}{23.8k}   &\multirow{2}{*}{57.1k}        & \multirow{2}{*}{772.9K}         \\ 
{\codevec}       &                       &                                &                               \\ \bottomrule \hline
\end{tabular}
}
\end{table}

%% file: sec/result.tex
\section{Result Analysis}
\label{sec:result}
\subsection{RQ1: Robustness of Neural Code Models}
\label{sec:rq1}
\input{table/rq1}

{\textbf{Approach.}} To answer this question, we re-use the original test sets of each model as a start point of the experiment. 
 Specifically, for each studied neural code model, we randomly selected 1,000 samples from its original test dataset.
 For each sample, we randomly select one of the ten mutation operators from Table~\ref{tab:opts} and apply the mutation operator to the sample to generate a test. 
 Note that, given a sample program, it is possible that some mutation operators are not applicable, e.g., \textit{Op4}-duplication cannot be applied if a program does not have any assignment statements. If this happens, we continue to randomly select another mutation operator until one mutated program is generated. {Based on our experiments, at least one of the ten mutation operators can be applied to any of the samples.} Hence, we have 1,000 mutated programs from the 1,000 samples for each neural model.
 Then we evaluate the three pre-trained neural models with the 1,000 sample (i.e., \textit{1k original}) and the generated 1,000 mutants (i.e., \textit{1k mutants}). 

{\textbf{Results.}} Table~\ref{tab:rq1} shows the impacts of the simple perturbations on the performance of the neural models. In particular, we show the performance (either BLEU or F1) of the studied neural code models under two test datasets, i.e., before and after perturbations. For all the three neural code models, we notice the performance on \textit{1k mutants} decreases comparing to \textit{1k original}. Specifically, for {\neuralcodesum}, the BLEU score reduces 69.5\% (from 40.82 to 12.46). The F1 scores of {\codeseq} and {\codevec} reduce 5.9\% (from 71.16\% to 66.96\%) and 4.44\% (from 47.68\% to 45.56\%) respectively. 

Although all the three neural code models are susceptible to semantic-equivalent transformations; however, the impact of simple perturbations on the performance differs across the three neural models.  
As we can see the performance of  {\neuralcodesum} has declined significantly compared to {\codeseq} and {\codevec}. The main reason is that {\neuralcodesum} uses one 
token sequence to represent 
{the entire body of a method} and most of our proposed mutation operators can introduce new tokens into the method body, which impacts the representation vector and may further impact the performance of the model. 
While {\codeseq} and {\codevec} use both AST paths and AST token information to represent the entire body of a method, which are more stable than the token-based representation of {\neuralCodeSum}. 
Thus these two models are less susceptible to the perturbation introduced by random mutation. 

\mybox{
{The studied neural code models face the robustness issues as their performance is negatively impacted by simple perturbations to test data. However, the negative impacts vary due to the use of different representations of code by each neural code model.}}

\subsection{RQ2: Comparison Across Different Mutation Operators}
\label{sec:rq2}
\input{figure/rq2_4_5_6}
\input{table/rq2_tab}
\input{table/rq2_nc}

{\textbf{Approach.}} To answer this question, we use the same \textit{1K original} test sets for the three neural code models collected in RQ1 (Section~\ref{sec:rq1}). 
For each mutation operator in Table~\ref{tab:opts},  
we apply it on the \textit{1K original} test dataset to generate new test data. 
In total, we generate 10 new test sets, i.e., each test set is generated by applying one particular mutation operator.
We then examine the performance of the three neural code models on each of the new test sets.

We further examine the differences in terms of neuron coverage between each new test set and the original test set.  
In particular, we compute the neuron coverage of each test data, then calculate the average neuron coverage of each test set. Also, we calculate the average number of newly activated neurons of each test set based on a pairwise comparison between an original test data and a mutated one, i.e., the activated neurons in a mutated data that are not activated by the original data. 

Moreover, we examine the difference between the two sets of activated neurons: one by an original test set, and the other by the mutated new test set. In particular, we calculate the Jaccard distance of each mutant sample against the original sample for each test set. 
Given two sets of neurons $N1$ and $N2$ respectively. We measure their Jaccard distance by $1-\frac{N1\cap N2}{N1\cup N2}$. 
The Jaccard distance can be a value between 0 and 1, with 1 indicating no overlap at all and 0 indicating a complete overlap between the two sets.

\textbf{Results.} \textit{\textbf{Comparison Across the Ten Mutation Operators}}.
Overall, from the results shown in Table~\ref{tab:rq2_comp}, we observe that the performance of each model decreases on each of the 10 new test sets, i.e., the performance decline of {\neuralcodesum} ranges from 9.16\% (\textit{Op10}) to 85.25\% (\textit{Op5}) and the performance decline of {\codeseq} and {\codevec} ranges from 0.14\% (\textit{Op10}) to 21.33\% (\textit{Op6}) and 0.23\% (\textit{Op2}) to 16.01\% (\textit{Op9}) respectively. 
In addition, we can also see that the most effective mutation operators for each model are different, e.g., \textit{Op5} is the most effective mutation operator for {\neuralcodesum}, while for {\codeseq} and {\codevec}, the most effective mutation operator is \textit{Op6} and \textit{Op9} respectively. 
We further conduct the Mann-Whitney U test ($p < 0.05$) to compare the performance of the three neural code models under test data with (i.e., \textit{Op1}--\textit{Op10}) and without perturbations (i.e., original test set). 
The results suggest that the performance decline caused by simple perturbations is statistically significant in all the ten test sets and for all the three studied neural code models.


Table~\ref{tab:rq2_nc} shows the comparison results across different mutation operators with regards to the impact on neuron coverage. 
Overall, the impact of each operator on neuron coverage varies across the studied neural code models. For example \textit{Op2}--\textit{Op7} significantly improve the neuron coverage on {\neuralCodeSum}, while decreasing the neuron coverage on {\codeseq}. This may be caused by the unique architecture of different models and properties of the test data. 
Interestingly, despite the decreased neuron coverage, we notice that all the mutation operators can activate new neurons compared to the original test sets. {This supports our design choice of using newly activated neurons instead of neuron coverage when guiding the search in fuzzing (line 12, Algorithm~\ref{alg:algorithm0})}.



Figure~\ref{fig:rq_2_ncs}, Figure~\ref{fig:rq_2_cs}, and Figure~\ref{fig:rq_2_nv} show the distribution of the Jaccard distance between the neuron coverage caused by each of the ten test datasets and the \textit{1k original} test data for {\neuralCodeSum}, {\codeseq}, and {\codevec} respectively. 
From these figures, we can see that the Jaccard distances vary for different operators and these results also confirm that different mutation operators activate different neurons at different rates.
\input{table/rq3}
\textit{\textbf{Op2--Op9 v.s. Op1 and Op10}}.
As we described in Section~\ref{sec:step1}, operators \textit{Op1} and \textit{Op10} are proposed and used in prior studies~\cite{yefet2019adversarial,zhang2020generating} to generate adversarial examples for neural code models.
Regarding the effectiveness on reducing the performance of a neural model for code, five (i.e., \textit{Op5}--\textit{Op8}) of the eight operators (i.e., \textit{Op2}--\textit{Op9}) can outperform \textit{Op1} and all can outperform \textit{Op10} on {\neuralCodeSum}. We can also observe similar results on {\codeseq} and {\codevec}, i.e., five and four of the new proposed eight mutation operators can outperform \textit{Op1} respectively and \textit{Op10} is worse than all the eight new operators. 
In this work, we use all the ten mutation operators for the mutation-based test case generation, as each operator represents a unique type of semantic preserving transformations, which may trigger a different part of the examined neural code models, their different neuron coverage also confirms this.

\mybox{All the ten mutation operators are shown to be effective in introducing perturbations that can significantly impact the performance of the studied neural code models. Our detailed analysis reveals that a mutated test set often activates a different set of neurons compared with the original test set. 
Last, the eight new mutation operators (i.e., \textit{Op2}--\textit{Op9}) are comparable or outperform the two operators (\textit{Op1} and \textit{Op10}) used in prior studies.}


\subsection{RQ3: Effectiveness of {\tool}}
\label{sec:rq3}

\noindent{\textbf{Approach.}} We reuse the \textit{1k original} test data collected in RQ1 (Section~\ref{sec:rq1}) to explore the performance of the NC-guided mutation generation in {\tool}. 
For each test program in \textit{1k original} dataset,  
we use {\tool} and the baseline approach \textit{Random@3} to generate new test data respectively. 
Note that, the number of new test data generated by {\tool} has an upper bound, i.e., three times of the seed programs (see details in Section~\ref{sec:overall}). 
The actual generated test set may contain less than the upper bound as the test data that does not activate new neurons is discarded. Meanwhile, \textit{Random@3} generates three new mutants for each seed program and yields a total of 3,000 generated programs. 
Thus the number of generated mutants of these two approaches might not be the same. 
Then we examine the performance of the three models on the two new test sets (i.e., {\tool} and \textit{Random@3}) respectively. 
Last, we calculate the average neuron coverage of each test set and the average Jaccard distance between each mutated test data and its original test data (i.e., one seed program in the \textit{1k original} test set). 


\noindent{\textbf{Results.}}
As we can see from the results in Table~\ref{tab:rq3}, both {\tool} and \textit{Random@3} can  generate new tests that detect more {classification} errors on the neural code models. With the newly generated tests, the performance of each examined model decreases significantly and the decline rate can be up to 84.81\% (i.e., NC-guided on {\neuralCodeSum}). 
In terms of neuron coverage, NC-guided strategy ({\tool}) achieves a lightly higher neuron coverage than Random@3 and \textit{1k original}. Compared with the 10 test sets by applying each mutation operator individually (Table~\ref{tab:rq2_nc}), the NC-guided strategy also achieves the highest neuron coverage. Based on Jaccard distance, on average, the NC-guide strategy generates a new test set that activates more neurons compared to \textit{Random@3}. 


\mybox{By utilizing coverage-guided fuzzing strategy, {\tool} is more  effective in testing neural code models than the baseline \textit{Random@3}, i.e., a lower BLEU or F1 value, a higher neuron coverage, and a higher ratio of newly activated neurons.}

\subsection{RQ4: Usefulness of {\tool} on Improving Models}

\label{sec:rq4}

\noindent{\textbf{Approach.}} Similar to Deeptest~\cite{tian2018deeptest}, as a proof-of-concept, we showcase the usefulness of the mutated test data by using a subset of the entire training data. In particular, we first train the three neural models with 10k randomly selected original training data from scratch. We reuse the validation dataset provided by each of the three models during the training process. 
Then for each neural code model, we apply {\tool} and \textit{Random@3} on the selected 10K training data to generate two new test sets. Combining the mutated test sets and the 10K randomly selected training data, we obtain two sets of enhanced training datasets, i.e., one enhanced by {\tool} and one by \textit{Random@3}. 
We then re-train each of the three models with the two enhanced training datasets respectively. 
Finally, we evaluate these re-trained models on the \textit{1K original} test dataset. 
\input{table/rq4}

\noindent{\textbf{Results.}} Table~\ref{tab:rq4} compares the performance across three types of training sets, i.e., original data, original data enhanced by Random@3 strategy, and original data enhanced by {\tool}. 
As we can see from the table, in all cases, the
performance of the re-trained model improved significantly over the original model and the improvements are 35.15\% on {\neuralCodeSum}, 8.83\% on {\codeseq}, and 34.14\% on {\codevec}.  


\mybox{Performance of the three neural code models can be improved 35.15\% on {\neuralCodeSum}, 8.83\% on {\codeseq}, and 34.14\% on {\codevec} by retraining the models with synthetic data generated by {\tool}.} 

%% file: table/rq1.tex
\begin{table}[t!]
\centering
\caption{ The results of testing the three neural code models on the test data before (i.e., \textit{1K original}) and after (\textit{1K mutants}) introducing simple perturbations.}
\label{tab:rq1}

\begin{tabular}{ccr}
\toprule \hline
\textbf{Model}         & \textbf{Test data} & \textbf{Performance (\%)} \\ \hline
\multirow{2}{*}{\neuralCodeSum} & \textit{1K original}   &  BLEU = 40.82 \\ 
           & \textit{1K mutants}    &  BLEU = 12.46    \\ \hline
\multirow{2}{*}{\codeseq}   & \textit{1K original}   &  F1 = 71.16  \\ 
           & \textit{1K mutants}    &  F1 = 66.96    \\ \hline
\multirow{2}{*}{\codevec}   & \textit{1K original}   &  F1 = 47.68    \\ 
           & \textit{1K mutants}    &  F1 = 45.56    \\ 
           \bottomrule \hline
\end{tabular}
\vspace{-0.1in}
\end{table}

%% file: figure/rq2_4_5_6.tex
\begin{figure*}[t!]
\centering
\begin{subfigure}{0.32\linewidth}
\centering
\includegraphics[width=0.85\textwidth]{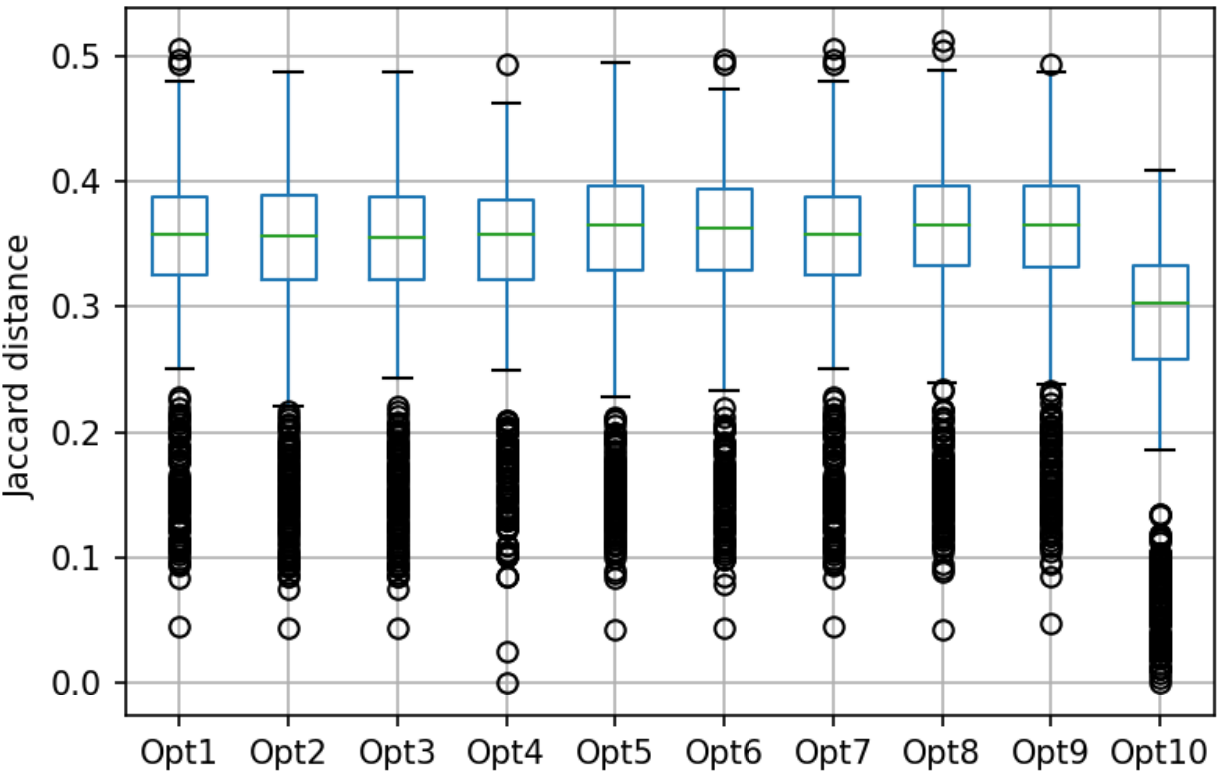}
\caption{{\neuralCodeSum}}
\label{fig:rq_2_ncs}
\end{subfigure}
\begin{subfigure}{0.32\linewidth}
	\centering
\includegraphics[width=0.85\textwidth]{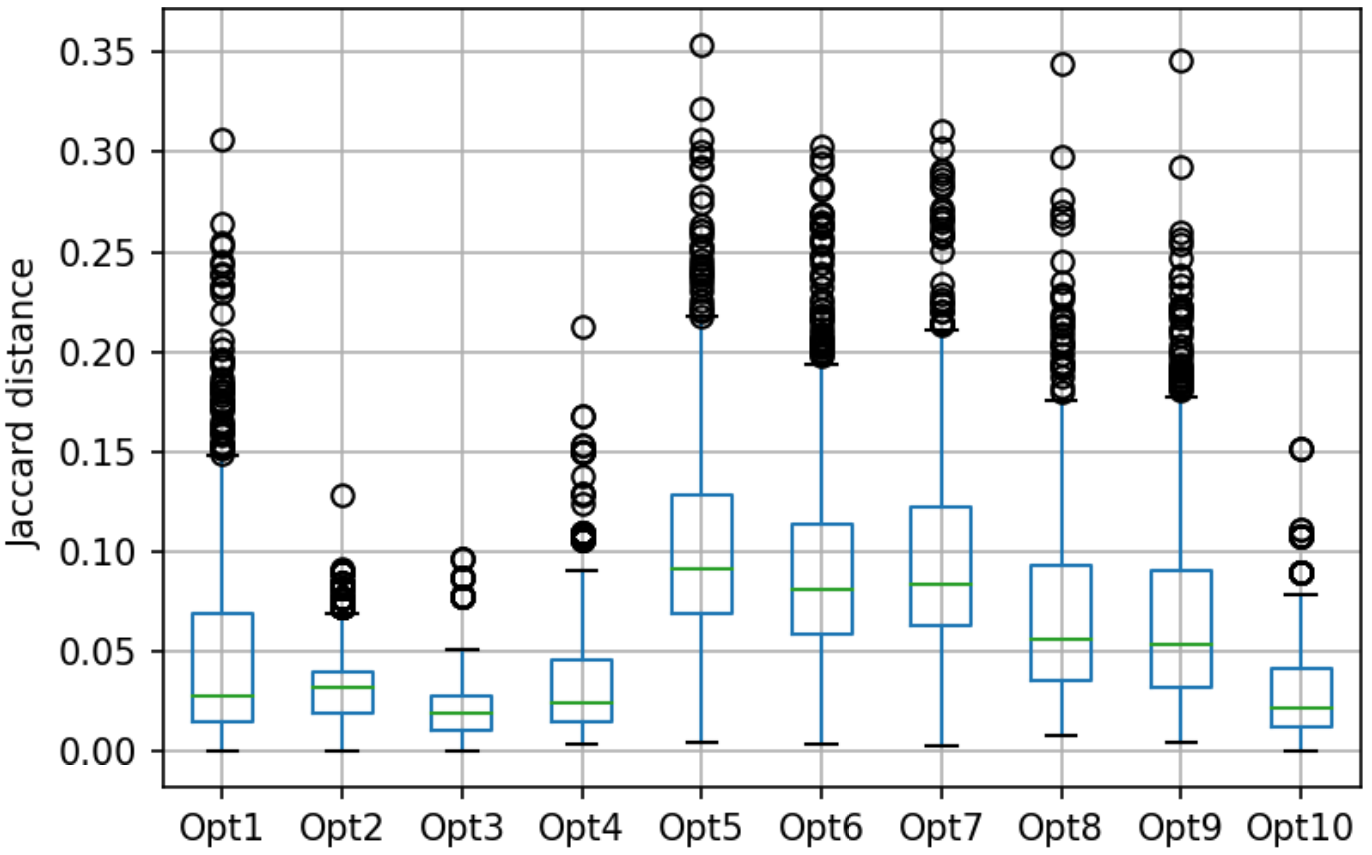}
	\caption{{\codeseq}}
\label{fig:rq_2_cs}
\end{subfigure}
\begin{subfigure}{0.32\linewidth}
	\centering
\includegraphics[width=0.85\textwidth]{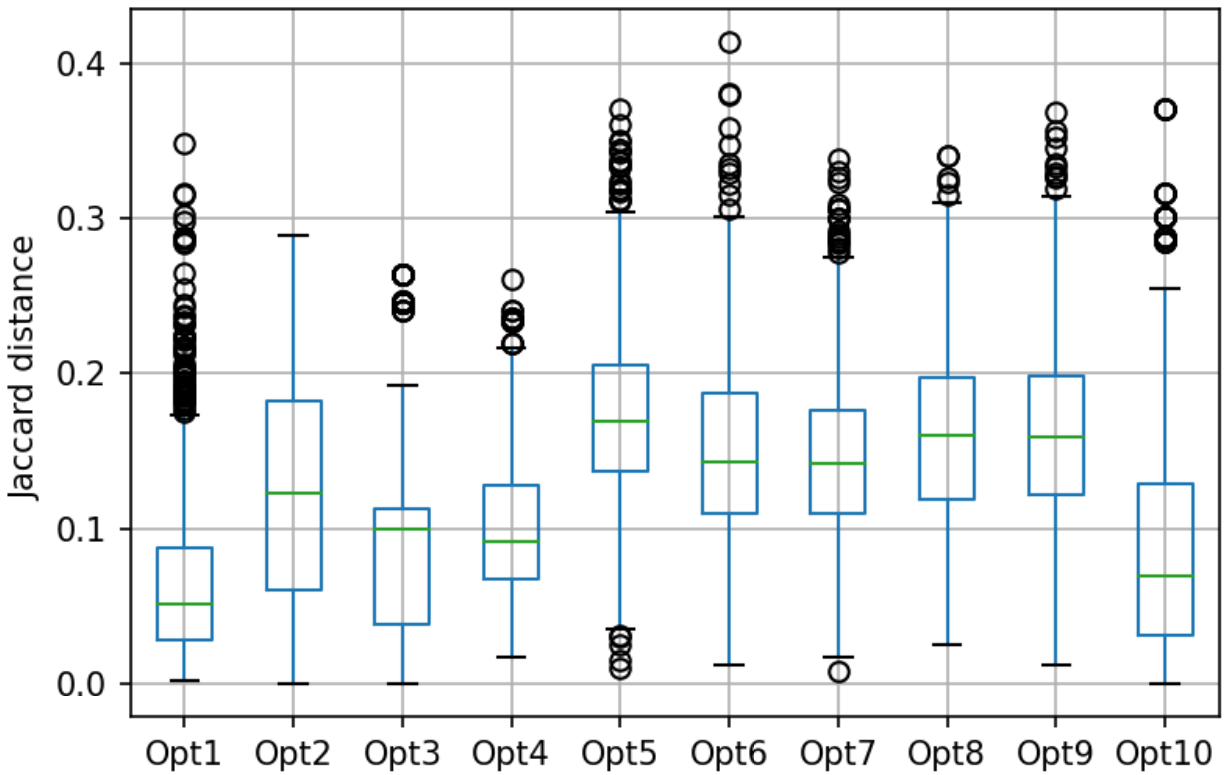}
	\caption{{\codevec}}
\label{fig:rq_2_nv}
\end{subfigure}\hfill
\vspace{-0.05in}
\caption{ 
Difference in neuron coverage caused by different mutation operators in the three models.}
\label{fig:rq2_4_5_6_plot}
\vspace{-0.1in}
\end{figure*}

%% file: table/rq2_tab.tex
\begin{table}[]

\caption{ Performance of different operators on {\neuralCodeSum}, {\codeseq}, and {\codevec}. 
`Original' shows the \textit{1k original} test set.  
`Op1'--`Op10' represent the test sets with the perturbations introduced by one mutation operator respectively.  
Numbers in the brackets are the performance decline of the examined neural models on the mutated test set compared with the original test set.}
\label{tab:rq2_comp}

\begin{tabular}{clll}
\toprule \hline
\multicolumn{1}{c}{}    & \multicolumn{1}{c}{\textbf{\neuralCodeSum}} & \multicolumn{1}{c}{\textbf{\codeseq}} & \multicolumn{1}{c}{\textbf{\codevec}} \\ 
& \multicolumn{1}{c}{BLEU (\%)}   & \multicolumn{1}{c}{F1 (\%)} & \multicolumn{1}{c}{F1 (\%)}     \\ \hline
\multicolumn{1}{l}{Original} & 40.82    &  71.16   &  47.68   \\\hline
\multicolumn{1}{l}{Op1}& 8.59 (78.95\%$\downarrow$)  &  68.23 (4.29\%$\downarrow$)  &  45.32 (5.05\%$\downarrow$) \\ 
\multicolumn{1}{l}{Op2}&  8.88 (78.26\%$\downarrow$)  &  70.75 (0.58\%$\downarrow$)   & 47.50 (0.23\%$\downarrow$)    \\ 
\multicolumn{1}{l}{Op3}&  8.84 (78.34\%$\downarrow$) &  70.70 (0.65\%$\downarrow$)   &    47.15 (0.97\%$\downarrow$) \\ 
\multicolumn{1}{l}{Op4}&  9.29 (77.21\%$\downarrow$)  & 70.85 (0.43\%$\downarrow$)    &   47.10 (1.08\%$\downarrow$)  \\ 
\multicolumn{1}{l}{Op5}&  \textbf{6.02 (85.25\%$\downarrow$)}  &  60.57 (17.49\%$\downarrow$)   &    44.73 (6.32\%$\downarrow$) \\ 
\multicolumn{1}{l}{Op6}&  6.16 (84.90\%$\downarrow$) &   \textbf{58.65 (21.33\%$\downarrow$)}  &  44.78 (6.32\%$\downarrow$) \\ 
\multicolumn{1}{l}{Op7}&  6.23 (84.73\%$\downarrow$)  &  59.82 (18.96\%$\downarrow$)   &    46.13 (3.21\%$\downarrow$)   \\ 
\multicolumn{1}{l}{Op8}&  7.19 (82.38\%$\downarrow$)  &  64.67 (10.04\%$\downarrow$)  &  43.81 (8.69\%$\downarrow$)  \\ 
\multicolumn{1}{l}{Op9}&  7.16  (82.45\%$\downarrow$) &  63.72 (11.68\%$\downarrow$)   & \textbf{41.04 (16.01\%$\downarrow$)}  \\ 
\multicolumn{1}{l}{Op10} & 37.08  (9.16\%$\downarrow$)   &    71.06 (0.14\%$\downarrow$) &  47.49 (0.25\%$\downarrow$)   \\ 
\bottomrule \hline
\end{tabular}
\end{table}


%% file: table/rq2_nc.tex
\begin{table}[t!]
\centering
\caption{ The average neuron coverage of the test sets generated using different mutation operators. Numbers in the brackets are the average number of newly activated neurons.}
\label{tab:rq2_nc}
\begin{tabular}{cccc}
\toprule \hline
   & \textbf{{\neuralCodeSum}} & \textbf{{\codeseq}} & \textbf{{\codevec}} \\ \hline
\multicolumn{1}{l}{ Original} &44.94\%&   90.79\% &  60.99\%\\ \hline
\multicolumn{1}{l}{Op1}&\textbf{47.33\%} (722.80)   &\textbf{91.46\%} (39.08)&  \textbf{61.48\%} (19.13)\\ 
\multicolumn{1}{l}{Op2}&\textbf{47.07\%} (702.65)         &   90.73\% (16.72) &  \textbf{61.00\%} (32.90)\\ 
\multicolumn{1}{l}{Op3}&\textbf{47.06\%} (702.32)         &   90.76\% (11.17) &  60.98\% (25.12)\\ 
\multicolumn{1}{l}{Op4}&\textbf{45.75\%} (700.31)         &   90.68\% (19.80)&  \textbf{61.17\%} (31.00)\\ 
\multicolumn{1}{l}{Op5}&\textbf{47.55\%} (745.05)         &   88.61\% (55.83)&  \textbf{62.37\%} (50.82)\\ 
\multicolumn{1}{l}{Op6}&\textbf{47.49\%} (742.95)         &   89.53\% (55.44) &  \textbf{62.33\%} (44.40)\\ 
\multicolumn{1}{l}{Op7}&\textbf{47.48\%} (741.40)         &   88.98\% (53.99) &  \textbf{62.04\%} (41.91)\\ 
\multicolumn{1}{l}{Op8}&\textbf{48.11\%} (768.31)&   \textbf{91.36\%} (52.39) &  \textbf{62.66\%} (48.44)\\ 
\multicolumn{1}{l}{Op9}&\textbf{48.10\%} (772.08)   &   \textbf{91.50\%} (53.21) &  \textbf{62.70\%} (49.41)\\ 
\multicolumn{1}{l}{Op10}&44.91\% (459.86)   &   \textbf{90.86\%} (21.82) &  60.97\% (25.17)\\ \bottomrule \hline
\end{tabular}
\vspace{-0.1in}
\end{table}

%% file: table/rq3.tex
\begin{table}[h]
\centering

\caption{Comparison results across original test sets, newly generated test sets using Random@3 and NC-guided ({\tool}) strategies. NC denotes Neuron Coverage and JD is the Jaccard Distance.}
\label{tab:rq3}
\setlength{\tabcolsep}{3pt}
\scalebox{0.9}{
\begin{tabular}{ccclcc}
\toprule \hline

\textbf{Model}   & \textbf{Strategies} & \textbf{\#New Tests} & \textbf{Performance (\%)} & \textbf{NC(\%)} &\textbf{JD} \\ \hline 
\multirow{3}{*}{\neuralCodeSum}     & \textit{1k original}    & -   &   BLEU = 40.82& 44.94 & - \\ 
                                & Random@3      &  3,000  &  BLEU=8.59 (78.95\%$\downarrow$) &  47.39      &   0.29    \\ 
                                & NC-guided     & 2,906   &   BLEU=6.20  (84.81\%$\downarrow$) &   48.95    &  0.32  \\ 
                             
                                \hline \hline
\multirow{3}{*}{\codeseq} & \textit{1k original}     & -   &   F1 = 71.16 & 90.79 &  - \\ 
                            & Random@3      &  3,000  &  F1=63.36 (10.96\%$\downarrow$) & 95.15    & 0.05 \\ 
                          & NC-guided     &  2,969  &  F1=55.46 (22.06\%$\downarrow$) & 95.71     & 0.08 \\ 
                          \hline \hline
\multirow{3}{*}{\codevec}  & \textit{1k original}     & -   &   F1 = 47.68 & 60.99    & - \\
                          & Random@3      &  3,000  &  F1=42.93 (9.96\%$\downarrow$)   & 72.24    &  0.17  \\ 
                          & NC-guided     &  3,000  &   F1=34.53 (27.58\%$\downarrow$) & 75.23    & 0.25   \\  
                                        
                          \bottomrule \hline 
\end{tabular}
}
\end{table}

%% file: table/rq4.tex
\begin{table}[t!]
\centering
\setlength{\tabcolsep}{3pt}
\caption{ \small Model retrain scenarios and the corresponding performance. 
\textbf{TrData} indicates the randomly selected 10K training data from the original training dataset. 
\textbf{TrData+{\tool}} indicates an enhanced training dataset by combining \textbf{TrData} and the synthetic inputs generated by applying {\tool} on \textbf{TrData}.
\textbf{TrData+Random} indicates an enhanced training dataset by combining \textbf{TrData} and the synthetic inputs generated by applying \texttt{Random@3} on \textbf{TrData};}
\label{tab:rq4}
\scalebox{1}{
\begin{tabular}{lll}
\hline
\textbf{Model}     & \textbf{Training data} & \textbf{\textbf{Performance (\%)}} \\ \hline
\multirow{3}{*}{\neuralCodeSum} &TrData & BLEU = 16.50  \\ 
   &TrData+Random & BLEU = 20.32   \\ 
   &TrData+{\tool}&  \textbf{BLEU = 22.30}    \\ \hline  \hline
\multirow{3}{*}{\codeseq}       &TrData & F1 = 22.98  \\  
   &TrData+Random &   F1 = 23.16    \\ 
   &TrData+{\tool} &  \textbf{F1 = 25.01} \\  \hline  \hline
\multirow{3}{*}{\codevec}       &TrData & F1 = 11.54  \\ 
   &TrData+Random  &  F1 = 15.13  \\ 
   &TrData+{\tool}      &  \textbf{F1 = 15.48}   \\ \hline
\end{tabular}
}
\end{table}

%% file: sec/discussion.tex
\section{Discussions}

\label{sec:discussion}

\subsection{Impact of the Applied Locations of Mutation Operators}
\label{sec:dis1}
Some of the mutation operators in {\tool} can be applied in any locations in a given program (i.e., location independent). For example, \textit{Op1} inserts an unused variable declaration into a randomly selected basic block of in a program, thus for any non-empty program there exists more than one location for \textit{Op1}. 
Among the ten mutation operators listed in Table~\ref{tab:opts}, \textit{Op1} and  \textit{Op5}--\textit{Op9} are location independent. 
To better understand the impact of this randomness on the performance of \tool, for each location independent operator, we apply it on the \textit{1K  original} test dataset to get a new test dataset. Then we collect the performance of the three models on the new test datasets. We rerun the above process 10 times and calculate the standard deviation values and the distribution of the performance of the three neural models. Table~\ref{tab:ds_1} shows the detailed results.  

We find that, for all the location-independent operators, their effectiveness on neural code model is insensitive to the applied locations, i.e., the variance is small. 
Our One-Way ANOVA test results show that there is no significant difference among the performance of the 10 runs, which suggests that the locations of generated mutants do not significantly impact the effectiveness of {\tool}. Thus, in our experiments we randomly pick a location to apply these mutation operators.
\input{table/ds_1}
\subsection{Distribution of the Selected Mutation Operators}
\label{sec:dist}
\input{table/ds_2}

{\tool} adopts a neuron coverage guidance algorithm to generate new tests with the ten pre-defined mutation operators. 
To further understand the operator selection process in {\tool}, we collected the selected operator for each mutant generated by {\tool} for each of the three neural code models, i.e., {\neuralCodeSum}, {\codeseq}, and {\codevec}. 
Table~\ref{tab:ds_2} shows the percentage of each operator used among the tests generated by {\tool} on each model. 
Overall, we can see that distribution of the selected operators varies dramatically among different neural code models, e.g., \textit{Op10} was used among 5\% of all the generated tests in {\neuralCodeSum} while less than 1\% of the generated tests from {\codeseq} used \textit{Op10}. 
In addition, we can also see that some of the operators are dominating across different models e.g, \textit{Op5} has been used in more than 15\% of the generated tests on each model. While these also exist {in the} operators that are selected less frequent across the three models, i.e., \textit{Op1},\textit{Op3}, \textit{Op4}, and \textit{Op10} are used in less than 10\% of the generated test cases across the three models. 
{One of the possible reasons is that \textit{Op5} activates relatively more new neurons than other operators (As shown in Table~\ref{tab:rq2_nc}). 
We have conducted Spearman rank correlation to compute the correlation between the number of newly activated neurons of mutants generated by an operator and the frequency of an operator used in a neural code model. The Spearman correlation values are 0.78 in {\neuralCodeSum}, 0.68 in {\codeseq}, and 0.92 in {\codevec}, which indicates that the frequency of an operator used in a neuron code model is positively correlated with its ability to activate new neurons (compared to the original tests).}

\subsection{Threats to Validity}
\label{sec:threats}
The selection of the studied neural code models and experimental projects could be a threat to validity. 
In this work, we only evaluated {\tool} on {\neuralCodeSum}, {\codeseq}, and {\codevec} with Java programs. 
Therefore, our results may not generalize to other neural code models or other programming languages. 
We leave the evaluation of the general applicability of our approach as future work. 
To generate new tests, this paper adopts ten different mutation operators for program transformations. 
Although these mutation operators have been examined can help generate effective mutants, these mutation operators may not represent many possible transformations for software programs. 
In addition, most of the used mutation operators are designed for object oriented programming languages, e.g., Java, which might not work for other program languages.
We will extend our approach with more mutation operators for supporting more program languages.  
{\tool} uses a threshold $MAX$ to limit the maximum number of mutation tries on a seed program. We set $MAX$ to three in this work {to simulate the natural statistics of unused code in software projects}, while the performance of {\tool} could vary with different values of $MAX$. We plan to explore the effectiveness of {\tool} with more $MAX$ values in the future. 
In this work, following existing studies~\cite{tian2018deeptest,xie2019deephunter} we use neuron coverage to guide the generation of valid mutants.  

In this work, following existing fuzzers for testing deep learning models, we also use neuron coverage to guide the generation of tests. Although there have been increasing discussions on whether neuron coverage is a meaningful metric, our experiments show that it works for testing neural code models. We plan to examine the effectiveness of {\tool} with more criteria. 

%% file: table/ds_1.tex
\begin{table}[t!]
\centering
\caption{\small Statistics of the impact of the position independent mutation operations used in this study.
\textbf{Performance Distribution} indicates the distribution of the performance of a model with 10 different test datasets generated by a mutation operator. 
\textbf{SD} is the standard deviation of the performance of a model with different test datasets. }
\label{tab:ds_1}
\begin{tabular}{lccc}
\hline
\textbf{Model}   & \textbf{Operator}   & \textbf{Performance Distribution} & \textbf{SD} \\
& & Average ($\pm$ Range) & \\\hline
\multirow{5}{*}{\begin{tabular}[c]{@{}l@{}}{\neuralCodeSum}\\ BLEU(\%)\end{tabular}}       & Op1& 7.10 ($\pm0.28$)     &  0.16   \\ 
 & Op5& 5.78 ($\pm0.43$)     &  0.27    \\ 
 & Op6& 5.88 ($\pm0.38$)     &  0.23   \\ 
 & Op7& 5.92 ($\pm0.40$)     &  0.24    \\ 
 & Op8& 6.50 ($\pm0.00$)     &  0.00     \\   
 & Op9& 6.61 ($\pm0.00$)     &  0.00        \\ \hline \hline
\multirow{5}{*}{\begin{tabular}[c]{@{}l@{}}{\codeseq}\\ F1(\%)\end{tabular}}            & Op1& 65.05 ($\pm1.47$)    &    0.01  \\ 
 & Op5& 63.98 ($\pm1.02$)    &    0.006  \\ 
 & Op6& 62.09 ($\pm1.56$)    &    0.009  \\  
 & Op7& 64.02 ($\pm1.26$)    &    0.008  \\ 
 & Op8& 64.65 ($\pm1.34$)    &    0.008  \\ 
 & Op9& 65.66 ($\pm1.14$)    &    0.008  \\   \hline \hline
\multirow{5}{*}{\begin{tabular}[c]{@{}l@{}}{\codevec}\\ F1(\%)\end{tabular}}            & Op1& 60.51 ($\pm2.46$)    &    0.02  \\
 & Op5& 59.95 ($\pm2.01$)    &   0.01  \\ 
 & Op6& 57.81 ($\pm2.44$)   &    0.01  \\
 & Op7& 60.34 ($\pm1.29$)   &    0.009  \\ 
 & Op8& 58.42 ($\pm2.14$)   &    0.01  \\  
 & Op9& 54.62 ($\pm1.91$)  &    0.01  \\ \hline
\end{tabular}
\vspace{-0.1in}
\end{table}

%% file: table/ds_2.tex
\begin{table}[t!]
\centering
\caption{ \small The distribution of selected mutation operators on the three models.}
\label{tab:ds_2}
\begin{tabular}{cccc}
\hline
\textbf{Operator} & \textbf{{\neuralCodeSum}} & \textbf{{\codeseq}} & \textbf{{\codevec}} \\ \hline
Op1     &    9.05\%                &    6.46\%           &     0.87\%     \\ 
Op2     &    8.39\%                &    10.83\%          &     15.30\%     \\ 
Op3     &    4.43\%                &    2.13\%           &     2.63\%     \\ 
Op4     &    2.82\%                &    0.63\%           &     1.23\%     \\ 
Op5     &    \textbf{15.79\%}      &    \textbf{19.00\%} &     \textbf{26.07\%}   \\ 
Op6     &    14.83\%               &    14.83\%          &     11.03\%    \\ 
Op7     &    13.97\%               &    9.13\%           &     5.53\%    \\ 
Op8     &    12.73\%               &    18.33\%          &     16.67\%   \\ 
Op9     &    13.21\%               &    16.50\%          &     19.37\%   \\ 
Op10    &    4.74\%                &    1.10\%          &     1.30\%     \\ \hline
\end{tabular}
\vspace{-0.1in}
\end{table}

%% file: sec/related.tex
\section{Related Work}
\label{sec:related}
\subsection{Testing Deep Learning Models}
\label{sec:rw_testDL}
In recent years, there are many studies on testing deep learning models~\cite{ma2018deepgauge,xie2019deephunter,tian2018deeptest,odena2019tensorfuzz,guo2018dlfuzz,pei2017deepxplore,wang2019adversarial,nejadgholi2019study}. 
Pei et al.~\cite{pei2017deepxplore} proposed DeepXplore to systematically find inputs that can trigger inconsistencies between multiple deep neural networks. 
They introduced neuron coverage as a systematic metric for measuring how much of the internal logic of a model have been tested. 
Tian et al.~\cite{tian2018deeptest} proposed DeepTest for failure detection on DNN-based Autonomous driving system. They adopted the neuron coverage metric as the criteria to generate synthetic inputs to test deep learning models. We use neuron coverage as guidance to our Mutation operator selection algorithm in our work. Ma et al.~\cite{ma2018deepgauge} purposed DeepGauge with 5 testing criterion for deep learning models, which extend the coverage metrics from neuron-level to layer-level. Odena et al.~\cite{odena2019tensorfuzz} presented TensorFuzz which applied fuzz-based coverage testing for deep learning systems. 
Nejadgholi~\cite{nejadgholi2019study} studied the oracle approximation issues in testing deep learning libraries. 
Guo et al. \cite{guo2018dlfuzz} proposed DLFuzz, a differential fuzzing testing framework to guide deep learning systems exposing incorrect behaviors. 
Xie et al.~\cite{xie2019deephunter} proposed DeepHunter, a fuzzing testing-based tool for testing deep learning models, in which they examined different types of seed selection strategy and test criteria.

The main difference between our work and the above studies is that most of the existing tools focus on general deep learning models in classic deep learning application domains, e.g., image processing, speech recognition, and natural language processing (NLP). While {\tool} is the first fuzzing framework for neural code models. 


\subsection{Adversarial machine learning.} 

Adversarial machine learning primarily focuses on generating adversarial examples to improve the performance of deep learning models~\cite{papernot2016limitations,cisse2017houdini,wang2019adversarial,gopinath2019symbolic,tuncali2018simulation,alcorn2019strike,alon2018general}. 
Gradient ascent-based adversarial example generation such as FGSM (Goodfellow et al. \cite{goodfellow2014explaining}) and BIM (Kurakin et al. \cite{kurakin2016adversarial}), which leverages the gradient of the model for finding adversarial example similar to the original input, has been widely used to accelerate the adversarial example generation problem for deep learning applications in the domains of image processing, speech recognition, and natural language processing. 

Recently, Yefet et al.~\cite{yefet2019adversarial} proposed DAMP, i.e., a gradient-based adversarial example generation technique, to generate adversarial examples for deep learning models in the domain of source code process. DAMP adopted two semantic preserving transformation operators, i.e., renaming variables and dead code inserting.  
Zhang et al.~\cite{zhang2020generating} proposed MHM that generated adversarial examples by renaming variables based on a sampling algorithm. Their experimental results demonstrate that MHM could effectively generate adversarial examples to attack the subject code process models. 
{Vahdat et al.~\cite{vahdat2021search} proposed a search-based testing framework for adversarial robustness testing. The differences to {\tool} are, firstly, out of the 10 operators {\tool} uses, there are 8 different operators compare to Vahdat et al.'s work. For the two similar operators (Renaming and Argument adding), {\tool} rename the variables with an 8-characters-length random string, while Vahdat et al.'s work uses the synonym of variable. Secondly, they used an evaluation metric derived from DeepMutation++~\cite{hu2019deepmutation++} to guide the mutation, while {\tool}  uses neuron coverage (NC) to guide the mutation.}



%% file: sec/conclusion.tex
\section{Conclusion}
\label{sec:conclusion}
This work proposes a coverage-based fuzzing framework, {\tool}, for testing deep learning-based models for code processing. In particular, we first propose and implement ten mutation operators to automatically generate valid (i.e., semantically preserving) source code examples as tests; 
we then propose a neuron coverage-based approach to guide the generation of tests.  
We investigate the performance of {\tool} on three state-of-the-art and typical neural code models,  i.e., {\neuralCodeSum}, {\codeseq}, and {\codevec}.  Our experiment results demonstrate that {\tool} can generate diverse and valid tests for examining the robustness and generalizability of neural code models. 
Moreover, the generated tests can be used to improve the performance of the target neural code models through adversarial retraining.